\documentstyle[prb,eqsecnum,aps]{revtex}
\font\bba=msbm10 scaled 1200
\font\bbb=msbm8 
\newfam\bbfam
\textfont\bbfam=\bba
\scriptfont\bbfam=\bbb
\def\bb{\fam\bbfam\bba}
\def\R{{\bb R}}

\begin{document}
\draft
\title
{ The density functional theory of classical fluids revisited.}
\author{J. M. Caillol \thanks{e-mail: caillol@labomath.univ-orleans.fr 
and 
caillol@lyre.th.u-psud.fr}}
\address{MAPMO - CNRS (UMR 6628)\\
         D\'epartement de Math\'ematiques \\
         Universit\'e d'Orl\'eans, BP 6759\\
         45067 Orl\'eans Cedex 2, France \\
	 and \\
	 LPT - CNRS (UMR 8627) \\
	 Bat. 210,
	 Universit\'e de Paris Sud \\
	 F-91405 Orsay Cedex, France}
         
\date{\today}
\maketitle
\begin{abstract}
We reconsider the density functional theory of nonuniform classical 
fluids from 
the point of view of convex analysis. From the observation that the 
logarithm of the grand-partition function $\log \Xi [\varphi]$ is a convex functional of 
the external potential $\varphi$ it is shown that  the Kohn-Sham free energy 
${\cal A}[\rho]$ is  a convex functional of the density $\rho$.  $\log 
\Xi [\varphi]$ and  ${\cal A}[\rho]$   constitute a pair of Legendre 
transforms and each of these functionals  can therefore be obtained as the 
solution of a variational principle. The convexity ensures 
the unicity of the solution in both cases. The variational principle 
which gives $\log \Xi [\varphi]$  as the maximum of a functional of 
$\rho$ is precisely that considered in the density functional theory 
while the dual principle, which gives  ${\cal A}[\rho]$ as the maximum 
of a functional of $\varphi$ seems to be a new result.  \\
KEY WORDS: Density fonctional theory; Theory of liquids; Convex analysis.
\end{abstract}
\pacs{}

\section{Introduction}
\label{intro}
The convexity properties of the thermodynamic potentials play an 
important role in  thermodynamics. For instance, 
in the case of a simple fluid, the entropy $S(U,V,N)$ is a concave function of
the internal energy $U$, the volume $V$ and the number of particles $N$. As
discussed  by Wightman 
\cite{Wightman} this
property, supplemented by the hypothesis that the entropy is a homogeneous
function of $(U,V,N)$ of the first degree (i.\ e. $S(\lambda U,\lambda V,\lambda
N)=\lambda S(U,V,N)$ for any $\lambda >0$), is fully 
equivalent to the second law of thermodynamics in Gibbs' 
formulation. \cite{Callen}
However, it
should be stressed that   the concavity of $S$ as a function  of
 $(U,V,N)$ - and, similarly,
the convexity of $U$ as a function of $(S,V,N)$- 
are properties which hold only in the thermodynamic limit where
the  properties of extensivity of $S$ and $U$ are valid. \cite{Wightman}
However,
there is one family of convexity properties which holds even in a finite volume.
That is the convexity of the logarithm of the canonical partition function 
$\log Z$ as a
function of the inverse temperature $\beta=1/kT$, and the convexity of the
logarithm of the grand partition function $\log \Xi$ 
in $\beta$ and $\nu=\beta \mu$ ($\mu$
chemical potential).\cite{Wightman} 

In this paper we extend the latter property to the
case of nonuniform systems. The system under interest is thus
a classical fluid in a volume ${\cal V}$ and in
the presence of an external one-body potential $\varphi(\vec{x})$. 
It will be  shown in Sec.\ III that $\log \Xi$ 
is a convex functional of the generalized potential
\begin{equation}
    \label{u}
u(\vec{x})=\nu -\beta \varphi(\vec{x}). 
\end{equation}
The implications of this mathematical property are studied in details
and are shown to lead to a new formulation of the density functional theory
(DFT) of classical fluids.
Indeed, the convexity of $\log \Xi\left[ u \right]$
implies the existence and the unicity of a
convex functional $\beta {\cal A} \left[ \rho \right] $
of the inhomogeneous density
$\rho \left( \vec{x} \right)$ such that
$\log \Xi\left[ u \right]$ and $\beta {\cal A} \left[ \rho \right] $ are
a couple of Legendre transforms.  Of course, the functional
$\beta {\cal A} \left[ \rho \right] $
identifies with the intrinsic Helmholtz (Kohn-Sham) free energy. 
Since they are Legendre transforms,
both $\log \Xi\left[ u \right]$ and $\beta {\cal A} \left[ \rho 
\right] $ can be expressed as the solution of a variational principle.
The variational
principle which defines $\log \Xi\left[ u \right]$ as the maximum of a functional
of $\rho(\vec{x})$ 
is precisely the one which was noticed by Kohn {\em et al.}
\cite{KohnI,KohnII}
and constitutes the cornerstone of the 
DFT.\cite{KohnI,KohnII,EvansI,EvansII,Oxtoby} The dual variational principle,
 which defines 
$\beta {\cal A} \left[ \rho \right] $ for a given $\rho \left( \vec{x} \right)$
as the maximum of a functional of $u(\vec{x})$ seems to have been 
overlooked.

Our paper is organized as follows. In Sec.\ II we review some mathematical
properties  of the grand-partition function  of a nonuniform fluid. 
In Sec.\ III we show that $\log \Xi\left[ u \right]$ is a convex 
functional of the generalized potential $u(\vec{x})$ which allows us 
to define its Legendre transform ${\cal A} \left[ \rho \right] $. The 
link with the DFT
is discussed in details. Conclusions are drawn in Sec.\ IV.
We have found usefull
to give  easy (formal) proofs of the
theorems of convex analysis which are needed in the text  in an
appendix. 
 
\section{ The Grand Partition function and its functional derivatives}
\label{I}
In this section we review known properties of nonuniform classical 
fluids and fix our notations.
We  consider a simple fluid made of identical, structureless classical particles
in a {\em finite} volume ${\cal V}$. The dimension of space is $d$.
A configuration of the system in
the grand-canonical ensemble will be denoted by $\omega \equiv (N,\vec{x}_1,
\ldots,
\vec{x}_N)$ where $\vec{x}_i$ denotes the position of the ith particle.
 The grand-canonical partition function $ \Xi\left[ u \right]$ can be written 
 as\cite{Hansen}
\begin{equation}
\Xi\left[ u \right]=
\int d \mu(\omega) \exp\left( \langle \hat{\rho}(\omega) \vert u \rangle
 \right) \; \label{def} ,
\end{equation}
where we have introduced the scalar product 
\begin{equation}
\langle \hat{\rho}(\omega) \vert u \rangle \equiv \int_{{\cal V}} d^d  \vec{x} \;
 \hat{\rho}(\vec{x} ;\omega)u(\vec{x}) \; 
\end{equation}
between the generalized potential  $u(\vec{x})$ (cf Eq.\ (\ref{u})) and and 
 the microscopic density of particles
$$\hat{\rho}(\vec{x} ;\omega)=\sum_{i=1}^N\delta^d(\vec{x}-\vec{x}_i) 
\; .$$ 
In Eq.\ (\ref{def}) we have adopted the compact notation
\begin{equation}
d \mu(\omega) = d \omega \exp \left(-\beta W(\omega)\right) \; , \label{dmu}
\end{equation}
where $W(\omega)$ denotes the internal potential energy of $N$ particles in the
configuration $\omega$ and $d \omega $ is the grand-canonical measure 
defined as 
\begin{equation}
\int d \omega \equiv \sum_{N=0}^{\infty} \frac{1}{\Lambda^{dN}N!}
 \int d^d \vec{x}_1 \ldots d^d \vec{x}_N  \;,
\end{equation}
where $\Lambda$ is  De Broglie thermal wave length. The grand-canonical
average of any microscopic variable ${\cal A}(\omega)$ can thus be 
written as
\begin{equation}
\langle {\cal A} \rangle=\int d \mu(\omega) \exp\left(
 \langle \hat{\rho}(\omega) \vert u \rangle
 \right) {\cal A}(\omega)/\; \Xi \left[u \right] \; .
 \end{equation}
We shall denote as usual by $\Omega[u]\equiv -\beta^{-1}\log \Xi [u]$ the grand 
Potential; note that it can be rewritten as a grand-canonical 
average  \cite{EvansI,EvansII,Oxtoby}
\begin{eqnarray}
    \label{logmean}
   \beta  \Omega[u] & =&  \langle \beta W(\omega) + \log f_{0}(\omega) 
    \rangle  -
    \langle 
    \rho \vert u \rangle \; , \\
    \rho (\vec{x})         &=& \langle \hat{\rho }(\vec{x};\omega) 
    \rangle \; ,
 \end{eqnarray}
 where $f_{0}(\omega)=\Xi^{-1} \; \exp \left(-\beta W(\omega) +  
 \langle \hat{\rho}(\omega) \vert u \rangle\right)$
is the grand-canonical equilibrium distribution in the phase space.

We shall denote by ${\cal U}$ the set of potentials $u(\vec{x})$ such that
 the r.h.s. of Eq.\
(\ref{def}) converges. For the sake of simplicity we shall restrict ourselves to
the case of $H$-stable systems in the sense of Ruelle, i.e. systems such that
$W(\omega) \geq -N {\cal B}$ where the lower bound ${\cal B<\infty}$ is some constant
independent of $N$.
Most models of interest satisfy this property. We have obviously
\begin{equation}
\Xi\left[ u \right]\leq \exp\left(\frac{1}{\Lambda^d} \exp\left(\beta {\cal
B}\right) \int_{{\cal V}} d^d \vec{x} \exp(u(\vec{x})) \right) \; , 
\end{equation}
from which it follows that ${\cal U}$ is the set of the functions $u(\vec{x})$
such that $\exp\left(u(\vec{x})\right)$ is integrable over the volume ${\cal V}$
of the system; i.e. 
$
{\cal U} = \left\{ u : {\cal V} \to \R \; ; \; \exp(u) \in L^1_{{\cal V}}
 \;[d^d \vec{x}]
\right\}\; $.
It can be shown that, for $H$- stable systems,  ${\cal U}$
is  a convex set.\cite{Chayes} Indeed,  let $u_0$ and $u_1$ two potentials
belonging to  ${\cal U}$. We  define the closed (respectively
open) interval $\left[u_0,u_1\right]$  as the set of functions
$u_{\lambda}=(1-\lambda)u_0 +\lambda u_1$ where $0\leq \lambda \leq 1 $
(respectively $0<\lambda<1$). As a consequence of the convexity of the
exponential function $\exp
\left( (1-\lambda)u_0 + \lambda u_1 \right) \leq (1-\lambda) \exp(u_0) 
+ \lambda \exp(u_1) $ and therefore $\left[ u_0,u_1 \right] \subset {\cal U}$,
QED.

As it is well known, $\Xi[u]$ is the generating functional of a hierarchy of
correlation functions. \cite{Hansen,Stell1,Stell2} Expanding the expression\ (\ref{def}) of 
$\Xi[u+\delta u]$ in powers of $\delta u$ one finds for $(u, u+\delta u \in {\cal
U})$

\begin{equation}
\Xi[u+\delta u]/\;\Xi[u]  =  \sum_{n=0}^{\infty} \frac{1}{n!}
\int_{{\cal V}} d^d\vec{x}_1 \cdots d^d\vec{x}_n \;  
G^{(n)} (\vec{x}_1,\ldots,\vec{x}_n) \; \delta u(\vec{x}_1) \cdots
 \delta u(\vec{x}_n) \; , 
\end{equation}
where the correlation function $G^{(n)}$ has been defined as
\begin{eqnarray}
G^{(n)} (\vec{x}_1,\ldots,\vec{x}_n) & = & \frac{1}{\Xi[u]} \frac 
{ \delta^{(n)} \; \Xi[u]} 
{\delta u(\vec{x}_1)\cdots  \delta u(\vec{x}_n)} \; \; \nonumber \\
&=& \langle \prod_{i=1}^n \hat{\rho}(\vec{x}_i;\omega)
\rangle \; .
\label{G}
\end{eqnarray}
It is sometimes also useful to introduce the connected correlation functions as
\begin{equation}
    \label{Gc}
G^{(n)}_c (\vec{x}_1,\ldots,\vec{x}_n)=
\frac { \delta^{(n)}\; \log \Xi[u]}{\delta u(\vec{x}_1)\cdots
  \delta u(\vec{x}_n)}
\; .
\end{equation}
The relation between $G^{(n)}_c$ and $G^{(n)}$ can be written symbolically
\begin{equation}
G^{(n)}_c (\vec{x}_1,\ldots,\vec{x}_n)= G^{(n)}(\vec{x}_1,\ldots,\vec{x}_n)
- \sum \prod_{m<n}€G^{(m)}_c (\vec{x}_{i_{1}},\ldots,\vec{x}_{i_{m}})\; ,
\end{equation}
where the sum of products is carried out over all possible partitions of 
the set $(1,\ldots,n)$ into subsets of cardinal $m<n$. 
\cite{Stell1,Stell2}
For instance we have 
\begin{equation}
G^{(2)}_c(\vec{x}_1,\vec{x}_2)=G^{(2)}(\vec{x}_1,\vec{x}_2)-\rho(\vec{x}_1)
\rho(\vec{x}_2) \; ,
\end{equation}
where  $\rho \equiv G^{(1)} \equiv G^{(1)}_c$
 is the mean density of particles. Note
that in the theory of liquids one rather defines the correlations functions $
\rho^{(n)}$ as the 
functional derivatives of $\Xi$ with respect to the activity $z=\exp(u)$. More
precisely : 
\begin{equation}
\rho^{(n)} (\vec{x}_1,\ldots,\vec{x}_n)= 
\frac{\prod_{1=1}^n z(\vec{x}_i)}{\Xi[z]}
\frac { \delta^{(n)}\; \log \Xi[z]}{\delta z(\vec{x}_1)\cdots
  \delta z(\vec{x}_n)}
\; .
\end{equation}
The
functions $\rho^{(n)}$ and $G^{(n)}$ differ by delta 
functions\cite{Hansen,Stell1}, for instance
$G^{(2)}(\vec{x}_1,\vec{x}_2)=\rho^{(2)}(\vec{x}_1,\vec{x}_2)+\rho(\vec{x}_1)
\delta^d(\vec{x}_1-\vec{x}_2)$.
\section{Convexity and Legendre transformations}
\subsection{Convexity of $\log \Xi[u]$}
\label{IIIA}
Let us show now that $\log\Xi[u]$ is a strictly convex functional of $u$, i.e.
that,
for any $u_1$ and $u_2 \neq u_1 \in {\cal U}$ we have, for any $0<\lambda<1$
the strict inequality
\begin{equation}
\label{convex}
\log \Xi[\lambda u_1 + (1-\lambda) u_2]<\lambda \log \Xi[u_1] +(1-\lambda) 
\log \Xi[u_2] \; .
\end{equation}
Let $u(\vec{x})$ be an arbitrary potential of ${\cal U}$ and $h(\vec{x})$ 
any non-zero function
defined on the volume of the system. The quadratic form
\begin{equation}
\left(\log \Xi \right) ^{(2)}_{u}[h]  \equiv \int_{{\cal V}} d^d \vec{x}_1  d^d \vec{x}_2 \;
\frac { \delta^{(2)}\; \log \Xi[u]}{\delta u(\vec{x}_1)
  \delta u(\vec{x}_2)} \; h(\vec{x}_1) h(\vec{x}_2) \;
\end{equation}  
will be shown to be strictly positive which implies the strict convexity from
 theorem\ II of  the appendix. Indeed it follows from Eqs.\ (\ref{G}) 
 and\ (\ref{Gc}) that
\begin{eqnarray}
\left(\log \Xi\right) ^{(2)}_{u} [h] &=&\int_{{\cal V}} d^d \vec{x}_1  d^d \vec{x}_2 \;
G_c^{(2)}(\vec{x}_1,\vec{x}_2)  h(\vec{x}_1) h(\vec{x}_2) \; , \nonumber \\
&=& \langle H^2 \rangle  - \langle H \rangle  ^2 >0 \; ,
\end{eqnarray}
where $H(\omega)=\sum_{i=1}^N h(\vec{x}_i)$.

 One can also give an alternative, more direct
proof of the convexity of $\log \Xi[u]$. For $u_1, u_2\neq u_1 \in {\cal U}$
and $0<\lambda<1$ we have, with slightly simplified notations,
\begin{eqnarray}
\label{Ho}
\Xi [\lambda u_1 +(1-\lambda)u_2] &=&
\int_{{\cal V}} d \mu \; 
\exp \left( \lambda \langle \hat{\rho} \vert u_1 \rangle \right)
\exp \left( (1-\lambda) \langle \hat{\rho} \vert u_2 \rangle \right)
\nonumber \\
&<& \left[\int_{{\cal V}} d \mu \; \left[
\exp \left( \lambda \langle \hat{\rho} \vert u_1 \rangle \right)
\right]^{1/\lambda} \right]^{\lambda}
\times 
\left[\int_{{\cal V}} d \mu \; \left[
\exp \left( (1-\lambda) \langle \hat{\rho} \vert u_2 \rangle \right)
\right]^{1/(1-\lambda)} \right]^{(1-\lambda)} \nonumber \\
&=& \Xi[u_1]^{\lambda} \; \Xi[u_2]^{(1-\lambda)} \; ,
\end{eqnarray}
 which  yields inequality\ (\ref{convex}) by taking the logarithm.
In order to write the second line of Eq.\ (\ref{Ho}) we have made use of
H\"older's inequality which states that, for 
any real numbers $p,q$ such that $p>1, \; q>1, \; 1/p +1/q =1 $ one 
has\cite{Tit}
\begin{equation}
    \label{hob}
    \vert \int d\mu(\omega) \; f(\omega) g(\omega) \vert \leq
    \left( \int d\mu(\omega)  \vert f(\omega) \vert^p \right)^{1/p} 
    \left( \int d\mu(\omega)  \vert g(\omega) \vert^q \right)^{1/q} 
    \; .
\end{equation} 
The case of equality in Eq.\  (\ref{hob}) occurs only if $\vert f(\omega) \vert^p /
\vert g(\omega) \vert^q$ is almost everywhere equal to a constant. 
This remark implies the strict inequality in Eq.\ (\ref{convex}).
Note that the grand-potential $\beta \Omega[u]=-\log \Xi[u]$
is therefore a
strictly concave functional of $u(\vec{x})$ even before the passage 
to the thermodynamic limit (TL). 
 \subsection{The Kohn-Sham free energy as the Legendre transform of $\log\Xi$}
 \label{IIIB}
Let us define the functional
\begin{equation}
\beta{\cal A}[\rho,u]=\langle \rho \vert u \rangle - \log \Xi[u].
\end{equation}
For a given function $\rho(\vec{x})$, $\beta{\cal A}[\rho,u]$ is
obviously a concave functional of $u$. Consequently, 
if $\beta{\cal A}[\rho,u]$ has
a maximum for some $\overline{u} \in {\cal U}$ this maximum is strict and
therefore unique. This is theorem\ III of the appendix. By definition, the value
of the maximum of $\beta {\cal A}[\rho,u]$  is the Legendre transformation
$\beta \overline{{\cal A}}[\rho]$ of the functional $\log \Xi [u]$ at
point $\rho$, i.e. we have
\begin{eqnarray}
\beta \overline{{\cal A}}[\rho] &=& \sup_{u\in {\cal U}} \beta{\cal A}[\rho,u]
\; ,  \label{a1}\\
&=& \beta{\cal A}[\rho,\overline{u}] \label{a2}\; ,
\end{eqnarray}
where $\overline{u}$ is the  solution of $\delta {\cal
A}[\rho,u]/\delta u(\vec{x})=0 $  for a given $\rho(\vec{x})$. If it 
exists,  $\overline{u}$ is therefore the unique solution of
\begin{equation}
\label{ubar}
\rho(\vec{x})=\left. \frac{\delta \log \Xi[u]}{\delta u(\vec{x})}
\right\vert_{u=\overline{u}} \; .
\end{equation}
As usual, existence is much more difficult to prove than unicity. Clearly, if for
some $\vec{x}\in {\cal V}$, $\rho(\vec{x}) <0$ Eq.\ (\ref{ubar}) will not have
any solutions in ${\cal U}$. It is easy to prove that, for the ideal gas
($W_N\equiv 0 \; \forall N$) the condition $\rho(\vec{x}) \geq 0 $ is sufficient
for Eq.\ (\ref{ubar}) to have a solution. For non-trivial models, this condition
is in general not sufficient since there exists an upper bound on
 $\parallel \rho
\parallel_{L^1}$. However, the beautiful result that for $H$- stable systems
the set ${\cal R}$ of densities for which Eq.\ (\ref{ubar}) have a solution is a
convex set has been proved  by Chayes and Chayes.\cite{Chayes}

Some important properties of the functional $\beta \overline{{\cal A}}[\rho]$
can easily be deduced from its mere definition. Firstly, 
 note Young's inequality which follows readily
from Eq.\ (\ref{a1})
\begin{equation}
(\forall u \in {\cal U}) \; (\forall \rho \in {\cal R}) \; \log \Xi[u] + \beta
\overline{{\cal A}}[\rho] \geq \langle \rho \vert u \rangle \; .
\end{equation}
Secondly, $\beta \overline{{\cal A}}$  is a strictly convex functional of $\rho$.
 To prove that, let $\rho_1,
\rho_2\neq \rho_1$ two distinct densities of ${\cal R}$ and $0<\lambda<1$. Since
${\cal R}$ is convex $\rho_{\lambda}=\lambda \rho_1 + (1-\lambda)\rho_2$ is also
a density of ${\cal R}$. Let $\overline{u}_{\lambda}$
 the solution of Eq.\ (\ref{ubar})
corresponding to $\rho_{\lambda}$. It follows from Eq.\ (\ref{a1}) that

\begin{eqnarray}
\beta \overline{{\cal A}}[\lambda \rho_1 +(1-\lambda)\rho_2] &=&
\lambda \beta {\cal A}[\overline{u}_{\lambda},\rho_1] +
(1-\lambda) \beta {\cal A}[\overline{u}_{\lambda},\rho_2] \; \nonumber \\
&<& \lambda \beta \overline{{\cal A}}[\rho_1] +(1-\lambda) 
\beta \overline{{\cal A}}[ \rho_2] \; ,
\end{eqnarray}
which proves the strict convexity.
An alternative proof of the convexity of $\beta \overline{{\cal A}}[ \rho]$ can
be found in Ref.\cite{Chayes}
Finally the functional derivative of $\beta \overline{{\cal A}}[ \rho]$
with respect to $\rho$ is easily obtained from Eq.\ (\ref{a2}). We have

\begin{eqnarray}
    \label{toto}
   \frac{\delta \beta \overline{{\cal A}}[\rho]}{\delta \rho(\vec{x}) }& =& 
 \frac{\delta \beta {\cal A}[\rho,\overline{u}]}{\delta 
 \rho(\vec{x}) }  \; , \nonumber \\
 &=& \overline{u}(\vec{x}) + \langle \rho \; \vert \;  \frac{\delta 
 \overline{u}}{\delta \rho(\vec{x})} \rangle 
 - \langle \frac {\delta \log \Xi}{\delta \overline{u}} \; \vert \;  \frac{\delta 
 \overline{u}}{\delta \rho(\vec{x})} \rangle \; , \nonumber \\
 &=&  \overline{u}(\vec{x}) \; ,
\end{eqnarray}
where we have made use of Eq.\ (\ref{ubar}).

Note that it follows from Eqs.\ (\ref{logmean}) and (\ref{a2}) that we 
can rewrite
\begin{equation}
    \label{KS}
    \beta \overline{{\cal A}}[\rho]=
    \langle \beta W(\omega) +\log \overline{f}_{0}(\omega) \rangle 
    \; ,
\end{equation}  
where $\overline{f}_{0}(\omega)$ denotes the grand canonical 
equilibrium density in the phase space for the system in the presence 
of the external potential $ \overline{u}$. The expression\ (\ref{KS}) 
of  $\beta \overline{{\cal A}}[\rho]$ coincides of course with that of 
the intrinsic Helmholtz (Kohn-Sham) free 
energy.\cite{EvansI,EvansII,Oxtoby}

We would like now to comment briefly on Eqs.\ (\ref{a1}), (\ref{a2}). 
They can be rewritten more explicitly as 
\begin{equation}
    \label{variaI}
    (\forall u(\vec{x}) \; ) \; \; 
  \beta \overline{{\cal A}}[\rho] \geq \int_{{\cal V}}d^d \vec{x} \;
  \rho(  \vec{x}) u(  \vec{x}) + \beta \Omega[u]  \equiv  
  \beta {\cal A}[\rho,u]  \; ,
\end{equation}   
where the equality holds for the unique potential $u\equiv
\overline{u}$ solution of Eq.\ (\ref{ubar}) . Eq.\ (\ref{variaI}) takes 
thus the form of a 
variational principle which asserts that the generalized potential 
$u(\vec{x})$ which gives rise to a given density $\rho(\vec{x})$ is that which 
maximises the functional $  \beta {\cal A}[\rho,u]$. If it exists it 
is unique. We have thus obtained a formal solution of the inverse 
problem to that usually considered in the DFT. To be effective this
variational principle requires an exact or at least an approximate 
knowledge of the functional $\beta \Omega[u]$. Unfortunately, although 
many efforts have been devoted in the past to find good approximations
of  $\beta \overline{{\cal A}}[\rho] $\cite{EvansI,EvansII,Oxtoby}
for many fluids of interest, similar approximates for
$\beta \Omega[u]$ are not available, at least to 
the author's knowledge.

The functional $\beta \overline{{\cal A}}[\rho]$ is the generating 
functional of the so-called direct correlation functions 
$\hat{c}^{(n)} (\vec{x}_1,\ldots,\vec{x}_n)$.\cite{Stell2}
These functions play in the theory of liquids
the role  devoted to  the vertex functions in statistical field 
theory.\cite{Zinn} They are defined as
\begin{equation}
\hat{c}^{(n)} (\vec{x}_1,\ldots,\vec{x}_n)  =  - \frac 
{ \delta^{(n)} \; \beta \overline{{\cal A}}[\rho]} 
{\delta \rho(\vec{x}_1)\cdots  \delta \rho(\vec{x}_n)} \; .
\label{chat}
\end{equation}
The direct correlation functions $\hat{c}^{(n)}$ and the Ursell 
correlation functions $G_{c}^{(n)}$ are related through generalized 
Ornstein-Zernike relations.\cite{Stell2,Zinn}
In fact, for  historical reasons, one rather defines the "true"
direct correlation functions $c^{(n)} (\vec{x}_1,\ldots,\vec{x}_n)$ by
\begin{equation}
\hat{c}^{(n)} (\vec{x}_1,\ldots,\vec{x}_n)  =  c^{(n)} (\vec{x}_1,\ldots,\vec{x}_n)  +
\hat{c}^{(n)}_{ id} (\vec{x}_1,\ldots,\vec{x}_n) ; ,
\end{equation}
where the $\hat{c}^{(n)}_{id}$ are the $\hat{c}^{(n)}$ 
functions of the ideal gas, so that the $c^{(n)}$ are the 
functional derivatives of minus the excess free energy. It is easily established that
\begin{eqnarray}
\hat{c}^{(1)}_{id} (\vec{x} ) &=& -\log\left(\Lambda^d 
\rho(\vec{x})\right) 
\; , \nonumber \\   
\hat{c}^{(n)}_{id} (\vec{x}_1,\ldots,\vec{x}_n)&=&
\frac{(-1)^{n-1}}{\rho(\vec{x}_1)^{n-1}} \; (n-2)! \prod_{i=2}^{n}
\delta ^d(\vec{x}_{1}-\vec{x}_{i}) \; (n\geq 2) \; ,
\end{eqnarray} 
so that, for instance we have  $ c^{(2)} (\vec{x}_1,\vec{x}_2) = 
\hat{c}^{(2)} (\vec{x}_1,\vec{x}_2)  +\delta^d( 
\vec{x}_1-\vec{x}_2)/\rho(\vec{x}_1)$.
Since the functional $\beta \overline{{\cal A}}[\rho]$ is convex, it 
follows by application of the theorem\ II of the appendix that the 
quadratic form $\beta \overline{{\cal A}}_{\rho}^{(2)}[h]$, where $h$ in an 
arbitrary no zero function defined on ${\cal V}$, is definite positive, which 
can be expressed as
\begin{equation}
    \int_{{\cal V}}d^d \vec{x}_1 d^d \vec{x}_2 \; \left(\frac{\delta^d( 
\vec{x}_1-\vec{x}_2)}{\rho(\vec{x}_1)}-c^{(2)} 
(\vec{x}_1,\vec{x}_2) \right)
h(\vec{x}_1) h(\vec{x}_2) \geq 0 \; .
\end{equation} 
Note that, in the above equation, {\em a priori} the inequality is 
non strict (see the appendix for more details on this point).
\subsection{The DFT recovered}
\label{IIIC}
As for functions of several variables \cite{Callen,Arnold,Crouzeix}, the 
Legendre transform of a functional is involutive, i. e., in our case, the 
Legendre transform of $\beta \overline{{\cal A}}[\rho] $ 
coincides with $\log \Xi [u]$. Indeed, since $\beta \overline{{\cal 
A}}[\rho] $ is a strictly convex functional defined on ${\cal R}$,
one can define its Legendre 
transform $\beta \overline{\Lambda}[u]$ for any $u \in{\cal U}$. One 
first defines the two fields functional 
\begin{equation}
    \beta \Lambda[u,\rho]  = \langle \rho \vert u \rangle -
  \beta \overline{{\cal A}}[\rho]  \; ,  \label{la0}
\end{equation} 
which is concave in $\rho$ for a given $u$. The Legendre transform 
$ \beta \overline{\Lambda}[u]$ is then defined as the maximum of 
$\beta \Lambda[u,\rho] $ at $u$ fixed, i.e.
\begin{eqnarray}
\beta \overline{\Lambda}[u] &=& \sup_{\rho\in {\cal R}} \beta 
\Lambda[u,\rho]
\; ,  \label{la1}\\
&=& \beta \Lambda[u,\overline{\rho}] \label{la2}\; ,
\end{eqnarray}
where $\overline{\rho}$ is solution of $\delta 
\Lambda [u,\rho]/\delta \rho(\vec{x})=0 $  for the given field  $u(\vec{x})$.  
If it exists, $\overline{\rho}$
is therefore the unique solution of 
\begin{equation}
\label{rhobar}
u(\vec{x})=\left. \frac{\delta \beta \overline{{\cal A}}[\rho] }{\delta 
\rho(\vec{x})}
\right\vert_{\rho=\overline{\rho}} \; .
\end{equation}
In order to prove that $\beta \overline{\Lambda} \equiv \log \Xi$ we insert the 
expression\  (\ref{a2}) of  $\beta \overline{{\cal A}}[\rho]$ in the 
r.h.s of Eq.\  (\ref{la2}) which yields
\begin{equation}
    \label{inter}
 \beta \overline{\Lambda}[u] = \langle \overline{\rho} \vert (u 
 -\overline{u}  ) \rangle + \log \Xi[\overline{u}] \; ,
 \end{equation}
where $ \overline{u}$ is the unique solution of 
\begin{equation}
    \overline{\rho}(\vec{x})=\left. \frac{\delta \log \Xi }{\delta 
u(\vec{x})}
\right\vert_{u=\overline{u}} \; .
\end{equation}
We claim now that $\overline{u}\equiv u$ and therefore that
$\beta \Lambda \equiv \log \Xi$.
Indeed it follows from Eq.\ (\ref{toto}) that  
$$
\overline{u}(\vec{x})=\left. \frac{\delta \beta \overline{{\cal A}}[\rho] }{\delta 
\rho(\vec{x})}
\right\vert_{\rho=\overline{\rho}} \;  ,  
$$
and hence, by comparison with Eq.\  (\ref{rhobar}) 
$\overline{u}\equiv u$, QED. 

We would like to comment briefly on Eqs.\ (\ref{la1}), (\ref{la2}). They 
can be recast under the more familiar form 
\begin{equation}
    \label{variaII}
    ( \forall \rho(\vec{x}) ) \; \; 
    \beta \Omega[u] \leq \beta \overline{{\cal A}}[\rho] -\int_{{\cal 
    V}}d^d \vec{x} \; \rho(\vec{x})  u(\vec{x}) \equiv \beta \Omega_{{\cal 
    V}}[u,\rho] ( \equiv -  \beta \Lambda[u,\rho]) \;, 
\end{equation}    
where the equality holds for the unique profile $\rho\equiv \ 
\overline{\rho}$  solution of Eq.\ (\ref{rhobar}). Eq.\ (\ref{variaII}) 
is precisely the DFT variational principle which states that the 
density $\rho(\vec{x}) $ corresponding to the imposed external 
potential $u(\vec{x}) $ is that which minimizes the functional 
$\beta \Omega_{{\cal V}} [\rho,u]$ for a given $u(\vec{x}) 
$.\cite{EvansI,EvansII,Oxtoby}
Moreover the solution is unique (if it exists).
\section{Summary and additional comments}

In this paper we have studied some of the properties of a nonuniform
classical fluid occupying a finite volume ${\cal V}$ and in the presence 
of an external potential $\varphi(\vec{x})$.
The thermodynamic and structural properties of the fluid are 
described by the logarithm of the grand partition function $\log \Xi 
[u]$ which is the generating functional of the connected correlation 
fonctions $G_{c}^{(n)}$.   $\log \Xi [u]$ is a convex functional of the 
generalized external potential $u(\vec{x})\equiv \beta \mu -\beta 
\varphi(\vec{x})$ and  its Legendre transform  
$\beta \overline{{\cal A}}[\rho] $, which identifies with the 
Kohn-Sham free energy, is a convex functional of the density 
$\rho(\vec{x})$. $\beta \overline{{\cal A}}[\rho] $ is  the generating 
functional of the direct correlation (or vertex) functions 
$\hat{c}^{(n)}$. The functions  $G_{c}^{(n)}$ and $\hat{c}^{(n)}$ are 
related by generalized Ornstein-Zernike relations. The 
Legendre transform is involutive and thus the Legendre transform of 
$\beta \overline{{\cal A}}[\rho] $ coincides with $\log \Xi [u]$. As 
a consequence of the convexity of $\beta \overline{{\cal A}}[\rho] $ 
and $\log \Xi [u]$ the two following quadratic forms are definite 
positive
\begin{eqnarray}
 \label{1}   \log \Xi ^{(2)}_{u}[h] & >& 0 \; , \\
 \label{2}    \beta \overline{{\cal A}}^{(2)}_{\rho}[h] & \geq & 0 \; .
\end{eqnarray}  
Moreover, for $H$-stable systems, the set ${\cal U}$ of the 
generalized external potentials $u$ and the set ${\cal R}$ of the densities $\rho$ 
are both convex sets. 

For arbirary potentials $u\in{\cal U}$ and densities $\rho \in {\cal R}$
we have Young inequality
\begin{equation}
\label{3}  
\beta \overline{{\cal A}}[\rho] + \log \Xi [u] \geq \langle \rho 
\vert u \rangle \; .
\end{equation}
$\beta \overline{{\cal A}}[\rho] $ and $\log \Xi [u]$ are both the unique 
solution of a variational principle :
\begin{eqnarray}
 \label{4} 
 \beta \overline{{\cal A}}[\rho] & =& \sup_{u \in {\cal U}} 
 \left( \langle \rho  \vert u \rangle - \log \Xi [u] \right) \; \; (\forall  
 \rho \in {\cal R}) \; ,  \\
 \label{5} 
 \log \Xi [u] &=& \sup_{\rho \in {\cal R}} 
 \left( \langle \rho  \vert u \rangle -  \beta \overline{{\cal A}}[\rho] \right)
\; \; (\forall    u \in {\cal U}) \; .
\end{eqnarray}
The latter variational principle\ (\ref{5}) is that of the DFT whereas 
the dual one\ (\ref{4}) is a new one.

All the results which are summarized above are valid for any finite 
domain ${\cal V}$, before the passage to the thermodynamic limit, and 
one can wonder whether they survive in the infinite volume limit. The 
notion of thermodynamic limit is not well defined for an 
inhomogeneous system and we concentrate henceforth on the case of homogeneous 
systems. The domain ${\cal V}$ that we consider now could be a cube of 
volume $V=L^d$ with periodic boundary conditions, or the surface of a 
hypersphere, etc. There is no applied external potential and $u=\beta \mu$ 
reduces to the chemical potential, so that our system is 
homogeneous even for a finite volume. The grand-canonical pressure $p_{V}$ is 
defined by the thermodynamic relation $\log \Xi (\mu)=\beta p_{V} V$, and 
similarly we define $a_{V} (\rho)=  \overline{{\cal A}}(\rho) /V$. Of 
course all the results of the paper apply to this pecular case and one 
concludes that, for a finite volume $V$,
the pressure $p_{V}(\mu)$ at given temperature and volume 
is a  {\em strictly } convex function of the chemical potential and 
that the specific grand-canonical free energy $ a_{V} (\rho)$ 
is a strictly convex function of the density. This is true even in the 
event of a liquid-vapor transition for instance. This behavior of 
 $a_{V} (\rho)$ contrasts with that of the canonical 
free energy $a_{V} ^{can}(\rho)$ which is a bimodal function of $\rho$ in the two phases 
region, due to surface effects. But, what happens in the thermodynamic limit?
We first note that the {\em strict} inequalities which enter 
the definition of strict convexity for both $p_{V}(\mu)$ and $a_{V}(\rho)$ can 
become equalities in the infinite volume limit. Therefore, in the TL, 
we can only claim that, at a given temperature,
$p_{\infty}(\mu)$ is a convex function of $\mu$ and 
that $a_{\infty}(\rho)$ is a convex function of the density.
Moreover, as it is well known, the continuity of some 
derivatives 
can also be lost in the TL. This is the case for $p_{\infty }(\mu)$ which 
exhibits a "cusp"  for some $\mu(T)$ at a temperature $T$ below the 
critical temperature $T_{c}$. One can however still define a Legendre 
transform in case of such discontinuities. Let us denote by $\rho_{g}$ 
and $\rho_{l}$ the two distinct values of $\partial p_{\infty } /\partial \mu$ at 
the cusp,
then $a_{\infty }(\rho)$ will be a linear function of $\rho$ in the interval $ 
[\rho_{g}, \rho_{l}]$. Cusps and intervalls correspond in a Legendre 
transformation. \cite{Wightman,Arnold} To summarize, in the coexistence region,
the {\em strict } convexity of  $a_{V}(\rho)$ 
is lost in the TL ;  by contrast,  $p_{\infty }(\mu)$ remains strictly 
convex in the TL even in the event of a phase transition, but its 
derivatives have discontinuity points.

\acknowledgments
I acknowledge F. James, J. J. Weis, J. L. Raimbault and D. Levesque
for useful discussions and comments. 

\appendix
\section{Convex functionals}
\label{a}
Functionals are a simple generalization of functions 
of several variables; they can be defined as "functions of 
functions". In this appendix we give formal proofs of the theorems on 
convex functionals \cite{Roka} which are needed in the text, without aiming at any
mathematical rigor.
The proofs given below are a mere transcription
of the material of any standard textbook on functions of several 
variables.\cite{Crouzeix}

We consider a convex set $X$ of real functions defined on a domain ${\cal V} 
\subset \R^d$ with a norm $\parallel \varphi \parallel  ( \varphi \in X) $.
We define a functional $ F$
as a mathematical object which 
associates to each element $\varphi$ a real number $F[\varphi]$.
$F$ will be said differentiable at point $\varphi$ of $X$ if it 
exits a linear functional $ F^{(1)}_{\varphi}$  such that
\begin{eqnarray}
    \label{dif1}
    F[ \varphi+ \delta \varphi] &
= &
 F[ \varphi] + F^{(1)}_{\varphi}[ \delta \varphi] +
\epsilon [\delta \varphi] \parallel \delta \varphi \parallel  \; , \\
\lim_{\parallel \delta \varphi \parallel \to 0} \epsilon [\delta \varphi]  
&=& 0 \; , 
\end{eqnarray} 
where we have noted
\begin{equation}
    F^{(1)}_{\varphi} [ \delta \varphi] =
    \int_{{\cal V}}d^d \vec{x} \; \frac{\delta F }{\delta 
    \varphi(\vec{x})} \; \delta \varphi(\vec{x}) \; ,
\end{equation}    
where $\delta F /\delta \varphi(\vec{x})$ is the functional 
derivative of $F$ with respect to the field $\varphi(\vec{x})$. One 
deduces from Eq.\ (\ref{dif1}) a useful expression for the 
derivative of $F$ in the direction $\delta \varphi$ :
\begin{equation}
    \label{deriv}
    F^{(1)}_{\varphi} [ \delta \varphi] = 
    \lim_{\lambda \to 0} \frac{ F[ \varphi+ \lambda \delta \varphi] - F[ \varphi]}
    {\lambda} \; \; (\forall \; \delta \varphi \in X ) \; ,
\end{equation}  
It is  easy to prove by applying  Rolle's theorem to the real 
function 
$g(t)=F(\varphi +t \delta \varphi)$ that aTaylor Mac-Laurin expansion 
also holds for functionals :
\begin{equation}
    F[ \varphi+ \delta \varphi] 
= 
 F[ \varphi] + F^{(1)}_{\varphi +\lambda \delta \varphi }[ \delta 
 \varphi] \; \; (0<\lambda<1) \; .
 \end{equation}
If $F$ is twice differentiable it generalizes as
\begin{equation}
    \label{mac2}
    F[ \varphi+ \delta \varphi] 
= 
 F[ \varphi] + F^{(1)}_{\varphi}[ \delta 
 \varphi] + \frac{1}{2} F^{(2)}_{\varphi +\lambda \delta \varphi }[ \delta 
 \varphi]
 \; \; (0<\lambda<1) \; ,
\end{equation}
where
\begin{equation}
    F^{(2)}_{\varphi }[ \delta  \varphi] = 
  \int_{{\cal V}}d^d \vec{x} \; d^d \vec{y} \; \frac{\delta ^{(2)}F }{\delta 
    \varphi(\vec{x}) \delta  \varphi(\vec{y})} \; \delta \varphi(\vec{x}) 
    \delta \varphi(\vec{y}) \; .
\end{equation} 
 It will be necessary to distinguish carefully between 
 and strictly 
 convex functionals. $F[\varphi]$ will be said convex if $(\forall 
 \varphi_{1}, \varphi_{2} \in X )$ and for all $0\leq \lambda \leq 1$ 
 one has $F[\lambda \varphi_{1} + (1-\lambda) \varphi_{2}] \leq 
\lambda F[ \varphi_{1}] +(1-\lambda) F[ \varphi_{2}]$, whereas it will 
be said strictly convex if $(\forall 
 \varphi_{1}, \varphi_{2}\neq \varphi_{1}  \in X )$ and for all 
 $0< \lambda < 1$ 
 one has $F[\lambda \varphi_{1} + (1-\lambda) \varphi_{2}] <
\lambda F[ \varphi_{1}] +(1-\lambda) F[ \varphi_{2}]$. $F[\varphi]$ 
will be said concave (resp. strictly concave) if $-F[\varphi]$ is 
convex (resp. strictly convex).
Henceforth we shall assume moreover that all the considered 
functionals are at least twice differentiable. 
Let us  establish our first theorem which states that a convex 
functional lies above its tangent plane, i.e. more precisely

{\em Theorem I} : 
\begin{itemize}
    \item $F$

    $\iff  (\forall \varphi_{1}, \varphi_{2} \in X ) 
    \; \;
    F[\varphi_{2} ] \geq F[\varphi_{1} ]+ 
    F^{(1)}_{\varphi_{1}}[\varphi_{2}- \varphi_{1}] \; . $
    
     \item $F$ strictly convex $\iff  (\forall 
     \varphi_{2} \neq  \varphi_{1} \in X ) 
    \; \;
    F[\varphi_{2} ] > F[\varphi_{1} ]+ 
    F^{(1)}_{\varphi_{1}}[\varphi_{2}- \varphi_{1}] \; . $
 \end{itemize}   
 Suppose $F$ convex and let $0<\lambda<1$ and 
 $(\varphi_{1},\varphi_{2} \neq  \varphi_{1} \in X ) $. The convexity 
 of $F$ implies that
 $$
 \frac{F[\varphi_{1}+\lambda (\varphi_{2}- \varphi_{1})] 
 -F[\varphi_{1}]}{\lambda} \leq F[\varphi_{2}]-F[\varphi_{1}] \; .
 $$
 Taking the limit $\lambda \to 0$ and making use of Eq.\ (\ref{deriv}) 
yields $ F[\varphi_{2} ] \geq F[\varphi_{1} ]+ 
 F^{(1)}_{\varphi_{1}}[\varphi_{2}- \varphi_{1}]$. If $F$ is 
strictly convex the above proof does not imply the strict  inequality, 
more care is needed. Let $0<\omega<1$. Since 
$$(1-\lambda)\varphi_{1} 
+\lambda \varphi_{2}= \frac{\omega -\lambda}{\omega} \varphi_{1} +
\frac{\lambda}{\omega} \left( \varphi_{1}+\omega 
(\varphi_{2}-\varphi_{1})\right)  \; ,$$ 
for $0\leq \lambda\leq \omega$ we have $0 \leq (\omega-\lambda)/\omega \leq 1$
and therefore
$$
F[\varphi_{1}+\lambda(\varphi_{2} - \varphi_{1})] \leq 
\frac{\omega - \lambda}{\omega} F[\varphi_{1}] +
\frac{ \lambda}{\omega} 
F[\varphi_{1}+\omega(\varphi_{2}-\varphi_{1})] \; .
$$
If $F$ is strictly convex one thus have for $0 < \lambda \leq \omega$
$$
\frac{F[\varphi_{1} + \lambda(\varphi_{2} - \varphi_{1})] - 
F[\varphi_{1}]}{\lambda} \leq \frac{F[\varphi_{1} + \omega (\varphi_{2} - \varphi_{1})] - 
F[\varphi_{1}]}{\omega} < F[\varphi_{2}] - F[\varphi_{1}]  \; .
$$
Taking the limit $\lambda \to 0$ and making use of Eq.\ (\ref{deriv}) 
yields the strict inequality : $ F[\varphi_{2} ] > F[\varphi_{1} ]+ 
    F^{(1)}_{\varphi_{1}}[\varphi_{2}- \varphi_{1}]  $,
since, by hypothesis, $ \omega < 1$.    
    
 Reciprocally, we assume that 
 $$
 (\forall \varphi_{1}, \varphi_{2} \in X ) 
    \; \;
    F[\varphi_{2} ] \geq F[\varphi_{1} ]+ 
    F^{(1)}_{\varphi_{1}}[\varphi_{2}- \varphi_{1}] \; . $$
 Let $\varphi_{1}$ and $\varphi_{2}$ two distinct points of $X$ and 
 $0<\lambda<1$, we have 
 \begin{eqnarray}
    F[\varphi_{1} ]  & \geq &  F[\lambda \varphi_{1}+(1-\lambda ) \varphi_{2}]  
    +  F^{(1)}_{\lambda \varphi_{1}+(1-\lambda ) 
    \varphi_{2}}[(1-\lambda) ( \varphi_{1}- \varphi_{2})] \; , \\
 F[\varphi_{2} ]  & \geq &  F[\lambda \varphi_{1}+(1-\lambda ) \varphi_{2}]  
    +  F^{(1)}_{\lambda \varphi_{1}+(1-\lambda ) 
    \varphi_{2}}[\lambda ( \varphi_{2}- \varphi_{1})] \; .   
 \end{eqnarray} 
 It is now sufficient to add the two above inequalities, multiplied 
 respectively by $\lambda$ and by $1-\lambda$ to obtain
 $$
 \lambda F[\varphi_{1}] + (1- \lambda) F[\varphi_{2}]  \geq F[\lambda 
 \varphi_{1}+(1- \lambda)\varphi_{2}] \; ,
 $$
 which establishes the convexity of $F$ or its strict convexity if the 
 inequalities are strict.

 Let us prove now the
 {\em Theorem II} : 
\begin{itemize}
    \item $F$ convex $\iff  (\forall \varphi_{1}, \varphi_{2} \in X ) 
    \; \;
    F^{(2)}_{\varphi_{1}}[\varphi_{2}- \varphi_{1}]  \geq 0 \; , $
    \item $(\forall \varphi_{1}, \varphi_{2} \neq \varphi_{1}  \in X ) 
    \; \;
   F^{(2)}_{\varphi_{1}}[\varphi_{2}- \varphi_{1}]  > 0 
   \Longrightarrow $  $F$ strictly convex.
 \end{itemize}   
 Let us first assume that $F$ is convex and let $\varphi_{1}$ an 
 arbitrary function of $X$. We define $G[\varphi]=F[\varphi] - 
 F^{(1)}_{\varphi_{1}}[\varphi - \varphi_{1}]$. It follows from the 
 convexity of $F$ and from theorem\ I that $ (\forall \varphi \in X) 
 \; G[\varphi] - G[\varphi_{1}] \geq 0$. We perform now a 2nd order 
 Taylor Mac-Laurin expansion of $G$ about $\varphi_{1}$ (cf. Eq.\ 
 (\ref{mac2}) ) which yields
 \begin{eqnarray}
     G[\varphi_{1}+t \varphi ] - G[\varphi] &=& \frac{t^{2}}{2} 
     \left( F^{(2)}_{\varphi_{1}}[\varphi] + \epsilon(t) \parallel 
     \varphi \parallel^2 \right) \; \geq 0 \; , \nonumber \\
     \lim_{t\to 0} \epsilon(t)&=&0 \; .
 \end{eqnarray}
Taking the limit $t\to 0$ yields $F^{(2)}_{\varphi_{1}}[\varphi]  \geq 
0 $. 

Reciprocally, let 
$\varphi_{1}$ and $\varphi_{2}$ ($\varphi_{2} \neq \varphi_{1}$) two 
arbitrary functions of $X$. It must exist some $ \varphi \in 
]\varphi_{1},\varphi_{2}[$ such that (cf. Eq.\ (A6))
\begin{equation}
\label{jh}
F[\varphi_{2}]=F[\varphi_{1}] + 
F^{(1)}_{\varphi_{1}}[\varphi_{2}-\varphi_{1}] +\frac{1}{2}
F^{(2)}_{\varphi}[\varphi_{2}-\varphi_{1}]  \; . 
\end{equation}
We define the (strictly) positive number $\rho $ by 
$\varphi_{2}-\varphi_{1}= \rho (\varphi-\varphi_{1})$ which allows us 
to rewrite\ (\ref{jh}) as 
$$ F[\varphi_{2}]-F[\varphi_{1}]  - 
F^{(1)}_{\varphi_{1}}[\varphi_{2}-\varphi_{1}] = \frac{\rho^{2}}{2}
F^{(2)}_{\varphi}[\varphi-\varphi_{1}] \; .
$$
Assuming the positiveness of the quadratic form of the r.h.s. yields, 
by application of theorem\ I, to the convexity of $F$ and to its
strict convexity in the case of strict positiveness.

The last result we need is {\em Theorem III} : 
\begin{itemize}
    \item If $F$ is convex and has a local minimum at  $\varphi_{1} \in X$, 
    then the minimum is global.
    \item If $F$ is strictly convex then $F$ has at most one minimum and 
    this is a strict  minimum.
\end{itemize}
Let $\varphi=\varphi_{1} + \delta \varphi $ an arbitrary function in 
$X$ and $0\leq \lambda \leq 1$. Since $F$ is convex
$$
F[\varphi_{1} +\lambda \delta \varphi] \leq (1 - \lambda) F 
[\varphi_{1}] + \lambda F[\varphi]  \; ; $$
hence $ F[\varphi_{1} +\lambda \delta \varphi] - F[\varphi_{1}] 
\leq  \lambda (F[\varphi] -F[\varphi_{1}] )$. Assuming the existence 
of a local minimum of $F$ at $\varphi_{1}$ implies that for some $0< 
\lambda_{0}<1$ we have $ F[\varphi_{1} +\lambda_{0} \delta \varphi] - 
F[\varphi_{1}] \geq 0$ from which it follows that 
$F[\varphi]-F[\varphi_{1}] \geq 0$. Therefore the minimum is global. 
If $F$ is strictly convex the same demonstration leads to the 
inequalities
$$
0 \leq  F[\varphi_{1} +\lambda_{0} \delta \varphi] -  
F[\varphi_{1}] < \lambda_{0} \left( F[\varphi]-F[\varphi_{1}]\right) 
\; , $$
which prove that the minimum is strict and therefore unique.


\end{document}